\documentclass[aps,prl,preprint,preprintnumbers,groupedaddress]{revtex4-1}
\usepackage{graphicx} % pdflatex
\usepackage{amsmath}
\usepackage{braket}
\usepackage{bm}
\usepackage{color}

\begin{document}
\preprint{NITEP 151}

%\preprint{}

\title{Microscopic study of the deformed neutron halo of $^{\bf 31}{\bf Ne}$} 

\author{R. Takatsu}
\affiliation{Department of Physics, Hokkaido University, Sapporo 060-0810, Japan}  
\author{Y.~Suzuki}
\affiliation{Research Center for Nuclear Physics (RCNP), Osaka University, Ibaraki 567-0047, Japan}
\author{W. Horiuchi}
\email{whoriuchi@omu.ac.jp}
\affiliation{Department of Physics, Osaka Metropolitan University, Osaka 558-8585, Japan}
\affiliation{Nambu Yoichiro Institute of Theoretical and Experimental Physics (NITEP),
Osaka Metropolitan University, Osaka 558-8585, Japan}
\affiliation{RIKEN Nishina Center, Wako 351-0198, Japan} 
\affiliation{Department of Physics, Hokkaido University, Sapporo 060-0810, Japan}  
\author{M. Kimura}
\email{masaaki.kimura@riken.jp}
\affiliation{RIKEN Nishina Center, Wako 351-0198, Japan} 
\affiliation{Department of Physics, Hokkaido University, Sapporo 060-0810, Japan}  

\date{\today}

\begin{abstract}
 The properties of the deformed halo of $^{31}{\rm Ne}$ were discussed using the antisymmetrized
 molecular dynamics plus resonating group method (AMD+RGM). The AMD+RGM calculations describe a
 large neutron radius for the ground state and reveal that the ground state is dominated by
 core-excited components. The resonant states were also investigated by applying the analytical
 continuation of  the coupling  constant. It was found that the first excited state (the $5/2^-$
 state) is also dominated by core-excited components and has a small decay width, whereas the second
 excited state (the $7/2^-$ state) is dominated by  the valence neutron in the $f$-wave coupled to
 the ground state of $^{30}{\rm Ne}$  
\end{abstract}

\maketitle

%#!pdflatex main.tex
\section{Introduction}

The formation of a neutron halo with a broad neutron distribution outside the core nucleus is a
unique phenomenon observed near the drip line~\cite{Tanihata1996}. Since their first discovery in 
$^{11}{\rm Li}$~\cite{Tanihata1985}, neutron haloes have been observed and discussed up to $sd$-$pf$
shell regions, such as $^{22}{\rm C}$, $^{29}{\rm F}$, and 
$^{37}{\rm Mg}$~\cite{Horiuchi2006,Tanaka2010,Horiuchi2012,Kobayashi2014,Takechi2014a,Bagchi2020}. 
$^{31}{\rm Ne}$ is one such example whose large nuclear radius has been reported from the measurement of
interaction~\cite{Takechi2012} and Coulomb breakup~\cite{Nakamura2009,Nakamura2014}
cross-sections. Note that in the normal shell ordering of spherical potentials, the valence neutron
of $^{31}{\rm Ne}$  occupies $f_{7/2}$; however, it must occupy the $s$- or $p$-wave to form the
neutron halo~\cite{Riisager1992}. Thus, the shell structure of $^{31}{\rm Ne}$ is modified from the normal one. 
This change in  shell order is ascribed to deformation. Hamamoto investigated the
properties of neutron orbits based on the Nilsson model (a deformed Woods-Saxon potential) to
explain the observed large Coulomb dissociation cross section, and noted that the valence neutron
likely occupies  particular orbits ([330\ 1/2], [321\ 3/2], or [200\ 1/2])~\cite{Hamamoto2010}. In
these cases, the ground state is either $J^\pi=3/2^-$ or $1/2^+$. This result is consistent with
shell model calculations~\cite{Poves1994,Nakamura2009,Nakamura2014} and particle-rotor model
calculations~\cite{Urata2011,Urata2012}; however, to  
understand the structure of the deformed halo, it is important to investigate this with a
full-microscopic model that can describe the halo structure.  

In a previous work~\cite{Minomo2011,Minomo2012,Sumi2012}, the antisymmetrized molecular dynamics
plus resonating group method 
(AMD+RGM) framework was applied to describe the spatially extended halo 
structure and was combined with a microscopic nuclear reaction model to discuss the large interaction
cross-section of $^{31}{\rm Ne}$. In this paper, we present the details of this structure
calculation and discuss the properties of the deformed halo of $^{31}{\rm Ne}$. We show
that the ground state of $^{31}{\rm Ne}$ is dominated by core-excited components, indicating the
strong coupling between the deformed halo and rotational excitation of the core nucleus. 
However, the AMD+RGM approach was inaccurate to describe resonance states. To overcome this
problem, in this study, we introduced the method of the analytical continuation of the coupling
constant (ACCC)~\cite{Kukulin1977,Tanaka1997}. We demonstrate that the combination of the AMD+RGM
and ACCC is successful in determining the energies and widths of resonances. The first excited
state, the $5/2^-_1$ state, is dominated by the core-excited components, whereas the second excited
state, the $7/2^-_1$ state, is not.  

This paper is organized as follows: In the next section, we explain the theoretical frameworks, that
is, the AMD+GCM, AMD+RGM, and ACCC, which are used to describe $^{31}{\rm Ne}$. In the
section~\ref{sec:result}, we compare the antisymmetrized molecular dynamics plus generator
coordinate method (AMD+GCM) and AMD+RGM results and discuss the deformed 
neutron halo. Based on the ACCC, we also discuss the excited resonant states built on top of the
deformed halo. The final section summarizes this study.

\section{Theoretical Framework}\label{sec:framework}
In this study, we applied the AMD+GCM and AMD+RGM frameworks to describe the neutron halo of
$^{31}{\rm Ne}$. The Hamiltonian is common in both frameworks and is given as
\begin{align}
 H = \sum_{i=1}^A t_i + \sum_{i<j}^A v_{ij} - t_{cm} ,
\end{align}
where the Gogny D1S interaction~\cite{berger1991} is used as effective nucleon-nucleon
and Coulomb interactions. The center-of-mass kinetic energy, $t_{cm}$, is exactly removed.

The variational wave function is also common; a parity-projected Slater determinant of
single-particle Gaussian wave packets are employed. 
\begin{align}
 \Phi^\pi_{\rm int} &=\frac{1+\pi P_x}{2}
  {\mathcal A} \set{\varphi_1\varphi_2\cdots\varphi_A},
 \label{eq:intwf}  
\end{align}
where $P_x$ is the parity operator. In this study, we focused on the positive-parity states
($\pi=+1$) of $^{30}{\rm Ne}$ and  the negative-parity states ($\pi=-1$) of $^{31}{\rm Ne}$.
A single-particle wave packet has a deformed Gaussian form~\cite{kimura2004},
\begin{align}
 \varphi_i({\bf r}) &= \prod_{\sigma=x,y,z}
 \exp\set{-\nu_\sigma(r_\sigma -Z_{i\sigma})^2}\chi_i\eta_i,
 \label{eq:singlewf} 
\end{align}
where $\chi_i$ is the spinor, and $\eta_i$ is the isospin fixed to either a proton or neutron.  
The parameters of the variational wave function are ${\bm Z}_i$, $\bm \nu$, and $\chi_i$. 
\subsection{AMD+GCM} 
The parameters of the variational wave function were optimized by the
energy variation with the constraint on the nuclear quadrupole deformation parameter $\beta$. After 
energy variation, we obtained the optimized wave functions $\Phi^{\pi}_{\rm int}(\beta)$ for  each
given value of $\beta$. The optimized wave functions were projected on the eigenstates of the
total angular momentum, 
\begin{align}
 \Phi^{J\pi}_{MK}(\beta) &= P^{J}_{MK}\Phi^{\pi}_{\rm int}(\beta) \nonumber \\
 &=\frac{2J+1}{8\pi^2} \int d\Omega D^{J*}_{MK}(\Omega) R(\Omega)\Phi^{\pi}_{\rm int}(\beta),
 \label{eq:prjwf} 
\end{align} 
where $P^{J}_{MK}$, $D^{J}_{MK}(\Omega)$, and ${R}(\Omega)$ denote the angular momentum projector,
Wigner $D$ function, and rotation operator, respectively. Then, we superposed the wave
functions with different values of the quadrupole deformation parameter $\beta$, 
\begin{align}
 \Psi^{J\pi}_{M\alpha} = \sum_{Ki} e_{Ki\alpha}\Phi^{J\pi}_{MK}(\beta_i).\label{eq:gcmwf} 
\end{align}
The coefficients $e_{Ki\alpha}$ can be obtained from the Hill-Wheeler equation~\cite{hill1953}, 
\begin{align}
 &\sum_{K'i'}(H_{KiK'i'}-E_\alpha N_{KiK'i'})e_{K'i'\alpha} = 0,\\
 &H_{KiK'i'} = \braket{\Phi^{J\pi}_{MK}(\beta_i)|H|\Phi^{J\pi}_{MK'}(\beta_{i'})}, \\
 &N_{KiK'i'} = \braket{\Phi^{J\pi}_{MK}(\beta_i)|\Phi^{J\pi}_{MK'}(\beta_{i'})},
\end{align}
where $E_\alpha$ is the eigenenergy of the eigenfunction given by Eq.~(\ref{eq:gcmwf}). 

This AMD+GCM framework has often been used for nuclear structure
calculations~\cite{Kanada-Enyo2003,Kanada-Enyo2012,Kimura2016}; however,
it has a disadvantage in describing neutron haloes because the asymptotics of the wave function are
restricted to a Gaussian form. To overcome this problem, we applied the AMD+RGM framework to $^{31}{\rm Ne}$.

\subsection{AMD+RGM and ACCC}
\begin{figure}[!h]
\centering\includegraphics[width=0.35\hsize]{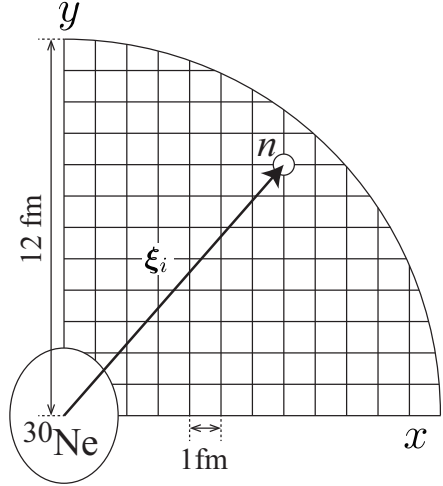}
\caption{Schematic illustration of an AMD+RGM basis wave function for the $^{30}{\rm Ne}+n$
 system.}  \label{fig:rgm} 
\end{figure}

In the AMD+RGM framework~\cite{Minomo2012,Dan2021}, we introduce an additional set of wave functions that
covers a large distance from the core nucleus. 
As schematically illustrated in Fig.~\ref{fig:rgm}, we constructed $^{30}{\rm Ne}+n$ wave functions,
which consist of $^{30}{\rm Ne}$ and a valence neutron located on a grid inside a 12 fm radius sphere
with 1 fm intervals,
\begin{align}
 \Phi(\beta,\bm \xi_i,\chi_n) =\mathcal{A}
  \set{\Phi_{^{30}{\rm Ne}}(\beta,\bm -\frac{1}{31}\bm \xi_i)\varphi_{n}(\frac{30}{31}\bm
 \xi_i,\chi_n)}, \label{eq:intrgm}
\end{align}
where $\Phi_{^{30}{\rm Ne}}$ is the intrinsic wave function of $^{30}{\rm Ne}$, obtained from the
energy variation with the constraint on $\beta$, and $\varphi_{n}$ is a Gaussian wave packet
[Eq.~(\ref{eq:singlewf})], which describes the valence neutron. 
To reduce the computational cost, 
we assumed axial and reflection symmetry for $\Phi_{^{30}{\rm Ne}}$. Hence, the relative
coordinate $\bm \xi_i$ between $^{30}{\rm Ne}$ and the valence neutron was restricted within the
first quadrant of the $xy$-plane, where the $y$-axis was the symmetry axis of $\Phi_{^{30}{\rm
Ne}}$. The generated basis functions were superposed with the basis functions obtained via the
variational calculations, 
\begin{align}
 \Psi^{J\pi}_{M\alpha} =& \sum_K\biggl\{\sum_{i=1}
 f_{Ki\alpha}\Phi^{J\pi}_{MK}(\beta_i) \nonumber \\
 & +\sum_{ij}\sum_{\chi_n=\uparrow,\downarrow}
 g_{Kij\chi_n\alpha}P^{J}_{MK}P^{\pi}\Phi(\beta_i,\bm \xi_j,\chi_n) \biggr\}. \label{eq:rgmwf} 
\end{align}
Similar to the AMD+GCM framework, the coefficients $f_{Ki\alpha}$ and $g_{Kij\chi_n\alpha}$ and
the eigenenergy were determined from the Hill-Wheeler equation. 

The AMD+RGM framework~\cite{Minomo2012,Dan2021} can describe a long-ranged halo wave function;
however, it is difficult to identify the resonances owing to the strong coupling with non-resonant
continua.  Therefore, we applied the ACCC~\cite{Kukulin1977,Tanaka1997} to calculate the energies
and widths of the resonances. We used the two-body spin-orbit interaction of the Gogny D1S
functional as an auxiliary potential,  
\begin{align}
 H(\lambda) &= H + \lambda v_{ls},\\
 v_{ls} &= iW_{ls}(\overrightarrow{\sigma_1}+\overrightarrow{\sigma_2})\cdot
 \overleftarrow\nabla\times\delta^3(\bm r_1-\bm r_2)\overrightarrow\nabla,
\end{align}
where $\lambda$ controls the strength of the auxiliary potential, and $\lambda = 0$ corresponds to
the physical point, that is, the original Hamiltonian. The eigenenergy of this Hamiltonian, which we
denote as $E(\lambda)$, varies as a function of $\lambda$. For larger values of $\lambda$, resonances
are bound owing to the additional attraction from the auxiliary potential, whereas non-resonant states
are insensitive to it. In such a way, we can distinguish   resonances from non-resonant
continua. The energy and width of a resonance can be calculated by the analytical continuation of the
eigenenergy $E(\lambda)$ from the bound region to the physical point. To this end, we introduced a
variable $X$, 
\begin{align}
 X = \sqrt{\lambda - \lambda_0},
\end{align}  
where $\lambda_0$ is the value at which $E(\lambda_0)$ is precisely zero. Using $X$, the wave number
of the valence neutron was approximated by a fractional function (the Pad\'e approximation),  
\begin{align}
 E(\lambda) &= \frac{\hbar^2}{2\mu}k^2(X) +E_{^{30}{\rm Ne}}(\lambda),\\
 k(X) &= i\frac{c_1 X +\cdots +c_MX^M}{1+d_1X+\cdots +d_NX^N}
\end{align}
where $\mu$ is the reduced mass of the two-body system, and $E_{^{30}{\rm Ne}}(\lambda)$ is the energy of
the ground state of $^{30}{\rm Ne}$, calculated from $H(\lambda)$. The $N+M$ coefficients were determined by
fitting the eigenvalue $E(\lambda)$ in the bound region. In this study, $N=M=7$ was found to
be sufficient for accurate approximation. Then, the wave number at the physical point,
$X_0=i\sqrt{\lambda_0}$, corresponds to the energy and width of  resonance as 
$\hbar^2k^2(X_0)/(2\mu) = E_R-i\Gamma_R/2$. 

\subsection{Valence neutron wave function}
To investigate the valence neutron wave functions, we calculate the overlap integral between the AMD+GCM
wave function of  $^{30}{\rm Ne}$ and the AMD+GCM or AMD+RGM wave function of $^{31}{\rm Ne}$,
\begin{align}
 \psi(\bm r) = \sqrt{31}\braket{\Psi^{J^{\pi}}_{M}(^{30}{\rm Ne})|
 \Psi^{J'^{\pi'}}_{M+m}(^{31}{\rm Ne})}. \label{eq:ofunc1} 
\end{align}
The computational method for this integral is explained in Ref.~\cite{kimura2017}. 
The multipole decomposition of Eq. (\ref{eq:ofunc1}) can be expressed as
\begin{align}
 \psi(\bm r)=\sum_{jl}C^{J'M+m}_{JMjm}u_{jl}(r)/r[Y_l(\hat r)\otimes \chi]_{jm},\label{eq:ofunc2} 
\end{align}
where $C^{J'M+m}_{JMjm}$ is the Clebsch-Gordan coefficient.
Here, $u_{jl}(r)$ is regarded as the valence neutron wave function coupled to the $^{30}{\rm Ne}$
core with a spin-parity of $J^\pi$. The squared integral of $u_{jl}(r)$ is the spectroscopic factor
($S$-factor),  
\begin{align}
 S_{jl} = \int_0^\infty dr\ |u_{jl}(r)|^2.
\end{align}

\section{Results and Discussion}\label{sec:result}
\subsection{AMD+GCM results}
\begin{figure}[tbp]
\includegraphics[width=0.5\hsize]{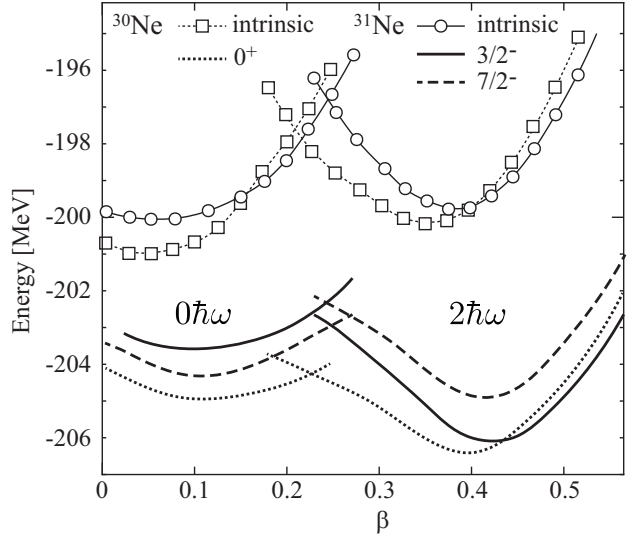}
 \caption{Energy curves of $^{30}{\rm Ne}$ and $^{31}{\rm Ne}$ obtained from $\beta$-constrained
 variational calculations.}
 \label{fig:surface}
\end{figure}
\begin{figure}[tbp]
\includegraphics[width=0.5\hsize]{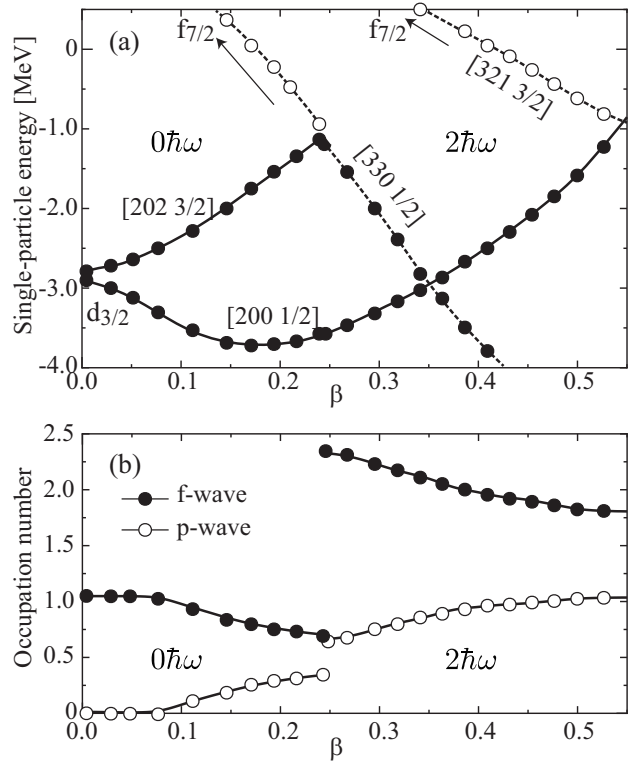}
 \caption{(a) Single-particle energies of the five most weakly bound neutrons. Only the occupied
 orbits are calculated in the AMD+GCM framework. (b) Occupation numbers of the $p$- and $f$-waves
 from the five most weakly bound neutrons.}  \label{fig:spl} 
\end{figure}
Figure~\ref{fig:surface} shows the energy curves of $^{30}{\rm Ne}$ and $^{31}{\rm Ne}$ obtained from
$\beta$-constrained variational calculations. Both nuclei exhibited
energy minima at approximately $\beta=0.1$ and 0.4. As shown later, these minima correspond to the
$0\hbar\omega$ and $2\hbar\omega$ configurations, respectively. The energies of the $0\hbar\omega$
and $2\hbar\omega$  configurations were inverted near $\beta=0.25$. To explain the structure of the
$0\hbar\omega$ and $2\hbar\omega$ configurations, Fig.~\ref{fig:spl}~(a) shows the single-particle
energies of the last five neutrons of $^{31}{\rm Ne}$.  In the $0\hbar\omega$ configuration ($\beta
<0.25$), the last neutron occupied the [330\ 1/2] Nilsson orbit originating from the spherical
$f_{7/2}$. In other words, the $0\hbar\omega$ configuration had one particle and zero holes (1p0h) with
respect to the $N=20$ shell closure. In the $2\hbar\omega$ configuration ($\beta > 0.25$), there were
two neutrons in the [330 1/2] orbit and one in the  [321\ 3/2] orbit, which is another intruder orbit
from $f_{7/2}$. Because the [202\ 3/2] orbit was unoccupied, the $2\hbar\omega$ configuration
corresponds to 3p2h. At larger deformations ($\beta > 0.6$), the 4p3h configuration appeared, in
which two neutrons occupied the [321\ 3/2] orbit and the last neutron occupied the [200\ 1/2] orbit,
although it is not shown because of its higher excitation energy. These three configurations (1p0h,
3p2h, and 4p3h) are in agreement with the analysis  by Hamamoto~\cite{Hamamoto2010}.

Note that these single-particle orbits are not the eigenstates of the orbital angular
momentum but mixed states due to  deformation. More specifically, the intruder orbits from the
spherical $f_{7/2}$ orbit ([330\ 1/2] and [321\ 3/2]) are the admixture of $f$- and $p$-waves. The
occupation numbers of the $f$- and $p$-waves in these intruder orbits are shown in
Fig~\ref{fig:spl}~(b). They discontinuously changed at $\beta=0.25$ because the
configuration switched from the $0\hbar\omega$ to $2\hbar\omega$. 
The sum of the occupation numbers of the $f$- and $p$-waves was approximately one or three,
corresponding to 1p0h and 3p2h.  If $\beta$ is close to zero, the last neutron mostly
occupies the spherical $f_{7/2}$ orbit, and hence, the occupation number of the $p$-wave was almost zero. As
deformation increased, the occupation number of the $p$-wave gradually increased. The ground state of
$^{31}{\rm Ne}$ approximately corresponds to $\beta=0.45$, where the occupation numbers of the $p$-
and $f$-waves are approximately one and two, respectively. 

After the angular momentum projection, the energy of the $2\hbar\omega$ minimum was lower than
that of the $0\hbar\omega$ minimum for both nuclei, indicating that the ground state is
dominated by the $2\hbar\omega$ configuration. Here, we focus on the energy curve of $^{31}{\rm Ne}$. 
For the $0\hbar\omega$ configuration, the $7/2^-$ state was lower than the $3/2^-$ state, indicating that
the spherical $f_{7/2}$ orbit is lower than the $p_{3/2}$ orbit in accordance with the normal shell
order. It is noted that the energy difference between the $7/2^-$ and $3/2^-$ states with the
$0\hbar\omega$ configuration was less than 1 MeV, which indicates the quenching of the spherical
$N=28$ shell gap. For the largely  deformed $2\hbar\omega$ configuration, the $3/2^-$ state was lower
than the $7/2^-$ state because the  valence neutron occupied the Nilsson orbit [321\ 3/2] and generated
rotational states with $K^\pi=3/2^-$. 

The comparison of the three configurations proposed by Hamamoto~\cite{Hamamoto2010} can be summarized as
follows: The 1p0h configuration ([330\ 1/2]): Our calculations did not yield a rotational spectrum
because of the small deformation and the spin-parity of the ground state is $7/2^-$, which cannot
have the halo structure. The 3p2h configuration ([321\ 3/2]): This yielded a rotational spectrum
with $K^\pi=3/2^-$ and can explain the halo structure of $^{31}{\rm Ne}$. The 4p3h configuration
([200\ 1/2]): The energy of this configuration has too large excitation energy and cannot be the
ground state if we attempt to consistently reproduce the properties of the neighboring nuclei. Thus,
3p2h is the primary candidate for the ground-state configuration. 

\begin{figure}[tbp]
\includegraphics[width=0.5\hsize]{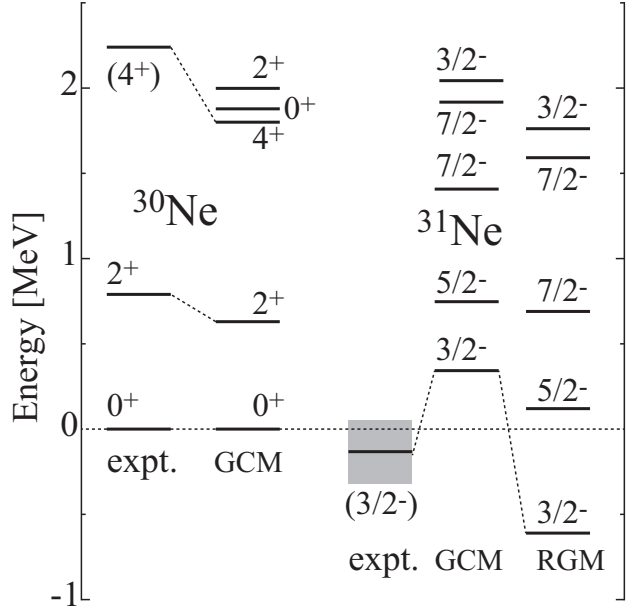}
 \caption{Energy spectrum of $^{30}{\rm Ne}$ and  $^{31}{\rm Ne}$ relative to the ground state of
 $^{30}{\rm Ne}$. The AMD+GCM results are denoted as `GCM' and the AMD+RGM results as `RGM'.} 
 \label{fig:level}
\end{figure}

% \begin{figure}[tbp]
% \includegraphics[width=0.9\hsize]{accc}
%  \caption{Tragectory of the energy eigenvalues of the $7/2^-$ states as function of the coupling
%  constant $\lambda$. The hatched area show the energies of widths of the $7/2_1$ and  $7/2_2$
%  resonances estimated by the ACCC.} 
%  \label{fig:accc}
% \end{figure}

Figure~\ref{fig:level} shows the spectra of $^{30}{\rm Ne}$ and $^{31}{\rm Ne}$ obtained from AMD+GCM. 
$^{30}{\rm Ne}$ was largely deformed and had the rotational ground band that consists of the $0^+_1$,
$2^+_1$, and $4^+_1$ states. Our calculation predicts the $0^+_2$ and $2^+_2$ states with $0\hbar\omega$
configuration at  approximately 2 MeV, although they have not yet been experimentally identified.
The ground state of  $^{31}{\rm Ne}$ was also dominated by the $2\hbar\omega$ configuration, which
exhibited a rotational spectrum with $K^\pi=3/2^-$ owing to large deformation. An obvious problem in
the AMD+GCM calculations is that $^{31}{\rm Ne}$ is unbound, and hence, we applied the AMD+RGM and
ACCC frameworks to improve the description of neutron haloes.

\subsection{AMD+RGM and ACCC results}
\begin{table}
 \caption{One-neutron separation energy in units of MeV ($S_n$), point-proton and neutron
 distribution radii in units of fm, and the quadrupole moments of protons ($Q_p$) and neutrons
 ($Q_n$) in units of $\rm fm^2$. The AMD+GCM results are denoted as `GCM' and the AMD+RGM results as `RGM'.} 
 \label{tab:gs}
\begin{ruledtabular}
\begin{tabular}{cccccc}
 &$S_n$ & $\sqrt{\braket{r_p^2}}$ & $\sqrt{\braket{r_n^2}}$ & $Q_p$ & $Q_n$ \\\hline
 $^{30}{\rm Ne}(\rm GCM)$  & --    & 3.05 & 3.40  & --   & --  \\
 $^{31}{\rm Ne}(\rm GCM)$  & $-0.36$ & 3.06 & 3.45  & 11.8 &  29.1\\
 $^{31}{\rm Ne}(\rm RGM)$  & 0.61  & 3.11 & 3.69  & 11.2 &  31.1\\
\end{tabular}
\end{ruledtabular}
\end{table}
Here, we show how the AMD+RGM calculation improves the description of the neutron halo and
bound the ground state of $^{31}{\rm Ne}$. As listed in Table~\ref{tab:gs}, the
calculated one-neutron separation energy ($S_n$) was 610 keV, which is approximately 1 MeV deeper
than the AMD+GCM result and slightly overestimated the experimental value of $150\pm 160$ keV deduced from
the one-neutron removal experiment~\cite{Nakamura2014}. The neutron distribution radius increased from 3.45
fm in the AMD+GCM to 3.69 fm in the AMD+RGM owing to the proper description of the neutron halo. The
proton radius also increased slightly owing to the recoil effect of the valence neutron. 
\begin{table*}
 \caption{One-neutron separation energy and widths in units of MeV, and the $S$-factors
 of the ground and excited states of $^{31}{\rm Ne}$ calculated using AMD+RGM and ACCC. The $S$-factors
 of the resonances  are calculated at the bound region close to $\lambda =
 \lambda_0$. The results of the AMD+GCM are also shown for the ground state.} 
 \label{tab:res}
\begin{ruledtabular}
\begin{tabular}{lcccccccc}
 & $S_n$ &$\Gamma$ & $0^+_1\otimes p_{3/2}$ & $2^+_1\otimes p_{3/2}$ & $4^+_1\otimes p_{3/2}$
 & $0^+_1\otimes f_{7/2}$ & $2^+_1\otimes f_{7/2}$ &  $4^+_1\otimes f_{7/2}$ \\\hline
 $3/2^-_1(\rm GCM)$  & $-0.36$ & --   & 0.13 & 0.32 & --   & --   & 0.78 & 0.66 \\
 $3/2^-_1(\rm RGM)$  & 0.61    & --   & 0.30 & 0.43 & --   & --   & 0.57 & 0.47 \\
 $5/2^-_1$           & $-0.12$ & 0.07 &  --  & 0.48 & 0.12 & --   & 0.66 & 0.33 \\
 $7/2^-_1$           & $-0.69$ & 0.18 &  --  & 0.07 & 0.21 & 0.76 & 0.05 & 0.06 \\
 & $S_n$ &$\Gamma$ & $0^+_2\otimes p_{3/2}$ & $2^+_2\otimes p_{3/2}$ & $0^+_2\otimes f_{7/2}$
 & $2^+_2\otimes f_{7/2}$ \\\hline 
 $3/2^-_2$  & $-1.76$  & 0.29 & 0.64 & 0.09 & --   & 0.23  \\
 $7/2^-_2$  & $-1.60$  & 0.21 & --   & 0.14 & 0.65 & 0.14
\end{tabular}
\end{ruledtabular}
\end{table*}

\begin{figure}[tbp]
\includegraphics[width=0.5\hsize]{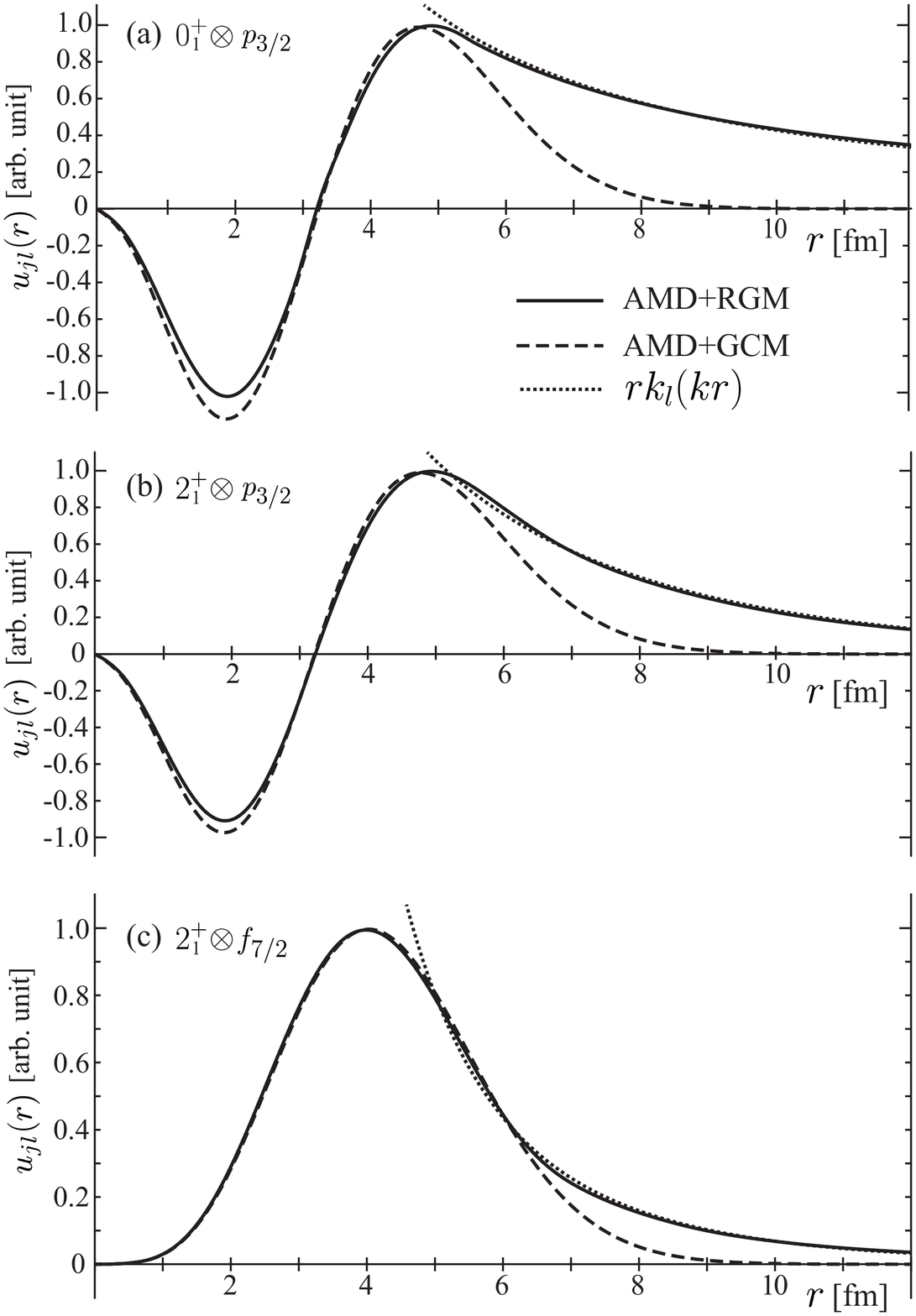}
 \caption{Wave functions (overlap functions) of the ground state of $^{31}{\rm Ne}$ in the (a)
 $0^+_1\otimes p_{3/2}$,  (b) $2^+_1\otimes p_{3/2}(2^+_1)$, and (c) $2^+_1\otimes f_{7/2}(2^+_1)$
 channels calculated using the  AMD+GCM and AMD+RGM. The amplitudes are arbitrarily normalized for 
 presentation. The dotted lines show the correct asymptotics at large distances, where the interaction
 between $^{30}{\rm Ne}$ and the valence neutron is negligible.}   
 \label{fig:ofunc}
\end{figure}

The large electric quadrupole moment predicted by the present calculations is a direct observable
to show core deformation. The quadrupole moment of neutrons was also large and slightly larger in
the AMD+RGM result than in the AMD+GCM result because of the development of the 
$p$-wave neutron halo with spatial anisotropy.  

To discuss the structure of the deformed halo in more detail, the spectroscopic factors of the
ground state are listed in Table~\ref{tab:res}. The main channels of the ground state are 
$J^\pi\otimes l_j=0^+_1\otimes p_{3/2}$, $2^+_1\otimes p_{3/2}$, $2^+_1\otimes f_{7/2}$, and
$4^+_1\otimes f_{7/2}$, where $J^\pi$ is the spin-parity of $^{30}{\rm Ne}$, and $l_j$ is the angular
momentum of the valence neutron. Compared with the AMD+GCM results, the
$S$-factors of the $p$-wave channels increased in the AMD+RGM results owing to the formation of a
$p$-wave halo.  
The one-neutron removal experiment  reported the $S$-factor of the
$0^+_1\otimes p_{3/2}$ channel as $0.32\pm0.21$~\cite{Nakamura2014}, which is reasonably reproduced by the
AMD+RGM calculation. It is remarkable that AMD+RGM calculations yielded larger $S$-factors for 
core-excited channels ($2^+_1\otimes p_{3/2}$, $2^+_1\otimes f_{7/2}$, and $4^+_1\otimes f_{7/2}$)
than the  $0^+_1\otimes p_{3/2}$ channel because of the strong coupling between the neutron halo and
the rotational excitation of the deformed core. This is qualitatively consistent with the results of 
the rigid-rotor model calculation~\cite{Urata2011}, which assumed large core deformation, and the
shell model calculation with the SDPF-M interaction~\cite{Nakamura2014}. In fact, the observed
inclusive parallel momentum distribution for the $1n$-removal reaction~\cite{Nakamura2014} shows
that the $f$-wave component is larger than the $p$-wave component, which is consistent with the
theoretical  calculations. Whether the reaction calculation can reproduce this momentum distribution
is an interesting problem for future study.

Figure~\ref{fig:ofunc} compares the wave functions of the valence neutron (overlap integral of
$^{31}{\rm Ne}$ and $^{30}{\rm Ne}$) in the $J^\pi\otimes l_j=0^+_1\otimes p_{3/2}$, 
$2^+_1\otimes p_{3/2}$, and $2^+_1\otimes f_{7/2}$ channels calculated using the AMD+GCM and
AMD+RGM. The AMD+RGM significantly improved the description of the wave function in
all channels and reproduced the correct asymptotics at large distances. The neutron wave function in the
$0^+_1\otimes p_{3/2}$ channel exhibited a very long tail due to the small $S_n$ and small orbital angular 
momentum, which is the origin of the large neutron halo. The  wave function of the
$2^+_1\otimes p_{3/2}$ channel also exhibited a similar distribution; however, it dumped faster than
that of the $0^+_1\otimes p_{3/2}$ channel because core excitation enlarged $S_n$. The
$2^+_1\otimes f_{7/2}$  channel had no halo tail owing to its large angular momentum. The wave
function of the $4^+_1\otimes f_{7/2}$ channel dumped as rapidly as that of the $2^+_1\otimes
f_{7/2}$ channel, although it is not shown in the figure.

The resonance energies and widths calculated using the ACCC are shown in Fig.~\ref{fig:level} and
Table~\ref{tab:res}. The $S$-factors  
of the resonances were calculated at the bound region close to $\lambda=\lambda_0$ because the
$S$-factors at the physical point cannot be calculated within the ACCC framework. We found that the
$S$-factors were not sensitive to the value of $\lambda$ in the vicinity of $\lambda=\lambda_0$;
hence we expect that these values are a reasonable approximation of the $S$-factors at the physical
point.

The $S$-factor shows us the structure of the resonances. The first excited state is the $5/2^-_1$
state for which the $0^+_1\otimes p_{3/2}$ channel is forbidden and the core-excited 
$2^+_1\otimes p_{3/2}$ and $2^+_1\otimes f_{7/2}$ channels dominate. Therefore, we regard this state as
the rotational excitation of the $^{30}{\rm Ne}$ core coupled to the neutron halo with  $p$- and
$f$-wave mixing. We also calculated the $S$-factor for the $0^+_1\otimes f_{5/2}$ channel and found it
to be very small. Note that this state is located above the $^{30}{\rm Ne}(0^+_1)+n$ threshold but below the
$^{30}{\rm Ne}(2^+_1)+n$ threshold; hence, the decay channel of the main component is
closed. Consequently, this state is a narrow and sharp resonance.  
The  $7/2^-_1$ state looks like a rotational state in the sequence of the $3/2^-_1$, $5/2^-_1$, and 
$7/2^-_1$ states. However, the $S$-factor indicates that this state is a single-particle state, that is,
it is dominated by the $0^+_1\otimes f_{7/2}$ component. Because the decay channel is open, its width
was larger than that of the $5/2^-_1$ state.  

Finally, we comment on the $3/2^-_2$ and $7/2^-_2$ states, which were significantly different from the
other states. They were strongly coupled to the spherical $^{30}{\rm Ne}(0^+_2)$ with the $0\hbar\omega$
configuration. Owing to small deformation, the coupling to $^{30}{\rm Ne}(2^+_2)$ was not strong. It
decayed through  weakly coupled $^{30}{\rm Ne}(0^+_1)$ channels, and the widths were broader due
to larger decay $Q$ values. The structure of this state was considerably different from the ground
state of $^{31}{\rm Ne}$ and the neighboring nuclei, indicating the coexistence of the deformed halo
state and spherical non-halo states. However, because of the mismatch of the structure, it would be
difficult to populate this state via the Coulomb excitation, neutron knockout, or transfer experiments.

\section{summary}\label{sec:summary}
In this study, we discussed the properties of the deformed halo of $^{31}{\rm Ne}$ by applying
the AMD+RGM and investigated the properties of the resonances built on top of the ground state
by applying the ACCC. By comparing the results from the conventional AMD+GCM framework, we showed
that the AMD+RGM correctly describes the asymptotic form of the neutron halo wave function and yields a
large neutron radius for $^{31}{\rm Ne}$. We also microscopically calculated the $S$-factors,
revealing that the ground state of $^{31}{\rm Ne}$ is dominated by  core-excited channels
($2^+\otimes p_{3/2}$ and $2^+\otimes f_{7/2}$) rather than the ground state channel. This result
explains the observed $S$-factor in the ground state channel, $0.32\pm 0.21$, reported by a
one-neutron removal experiment~\cite{Nakamura2014}. Simultaneously, it shows that $^{31}{\rm Ne}$ is
a deformed halo nucleus in which the rotational excitation of the core is coupled with a deformed
neutron halo.  

Furthermore, by applying the ACCC to the AMD+RGM, the energy and width of the resonances were
obtained. The first excited state, the $5/2^-$ state, was dominated by core-excited
components. Because  neutron decay in the dominant channel is a negative Q-value, this state had a 
small width. In contrast, the second excited state, the $7/2^-$ state, is dominated by the $f_{7/2}$
channel. Other resonances were  predicted to have the valence neutron coupled to the
spherical $0^+_2$ state of $^{30}{\rm Ne}$. 

We note that the successful combination of the AMD+RGM and ACCC frameworks will allow us to discuss
the properties of other deformed haloes, such as $^{37}{\rm Mg}$, and the nuclei beyond the neutron
drip line, such as $^{26}{\rm O}$.

\begin{acknowledgments}
This work was supported by the the collaborative research program 2022 at Hokkaido University, and
JSPS KAKENHI Grant Nos.	18K03635,  19K03859, 21H00113, and 22H01214.
\end{acknowledgments}

\bibliography{main}
\end{document}